\documentclass[traditabstract]{aa} 
\usepackage{graphicx,natbib,longtable,txfonts,lscape}
\usepackage[switch]{lineno}

\newcommand{\kms}{$\rm{\,km \,s}^{-1}$}

\newcommand{\FR}{FRI{\sl{CAT}}}
\newcommand{\FRo}{FR0{\sl{CAT}}}
\newcommand{\sFR}{sFRI{\sl{CAT}}}
\newcommand{\Ndue}{$N_{\rm cn}^{2000}$}
\newcommand{\Nuno}{$N_{\rm cn}^{1000}$}
\newcommand{\Nduemed}{$\bar{N}_{\rm cn}^{2000}$}
\newcommand{\Rcn}{$d^{\rm cn}_{\rm proj}$}
\newcommand{\vpar}{$c\Delta z$}

\usepackage{color}

\begin{document}
\defcitealias{baldi18h}{BCM18}
\defcitealias{massaro19h}{M19}
\setlength{\linenumbersep}{3pt}
   \title{The large-scale environment of FR~0 radio galaxies.}

   \subtitle{}

   \author{
A. Capetti\inst{1}
          \and
          F. Massaro\inst{2,1,3,4}
\and
          R.~D. Baldi\inst{2,5}
          }

   \institute{INAF-Osservatorio Astrofisico di Torino, via
     Osservatorio 20, 10025 Pino Torinese, Italy, \and Dipartimento di
     Fisica, Universit\`a degli Studi di Torino, via Pietro Giuria 1,
     10125 Torino, Italy, \and Istituto Nazionale di Fisica Nucleare,
     Sezione di Torino, I- 10125 Torino, Italy, \and Consorzio
     Interuniversitario per la Fisica Spaziale, via Pietro Giuria 1,
     I-10125 Torino, Italy, \and Department of Physics and Astronomy,
     University of Southampton, Highfield, SO17 1BJ, UK} \date{}

  \abstract {We explore the properties of the large-scale environment
    of FR~0 radio galaxies belonging to the FR0{\sl{CAT}} sample which
    includes 104 compact radio sources associated with nearby
    ($z<0.05$) early-type galaxies. By using various estimators we
    find that FR~0s live in regions of higher than the average galaxies
    density and a factor two lower density, on average, with respect
    to FR~I radio galaxies. This latter difference is driven by the
    large fraction (63\%) of FR~0s located in groups formed by less
    than 15 galaxies, an environment which FR~Is rarely (17\%)
    inhabit. Beside the lack of substantial extended radio emission defining the
    FR~0s class, this is the first significant difference between the
    properties of these two populations of low power radio galaxies.
    We interpret the differences in environment between FR~0s and
    FR~Is as the due to an evolutionary link between local galaxies
    density, BH spin, jet power, and extended radio emission.}

\maketitle

\section{Introduction}

The connection between the large scale environment and the
  properties of extra-galactic radio sources has been explored since
  the '70s (e.g., \citealt{longair79}). \citeauthor{longair79} found
  that weak radio galaxies (RGs) show no tendency to belong to groups
  or clusters of galaxies, an environment which is instead typical of
  extended powerful radio sources. \citet{prestage88} claimed that
  Fanaroff-Riley class II sources \citep{fanaroff74}, FR~IIs, as well
  as compact radio sources, lie on average in poorer clusters than
  those of class I (FR~Is). \citet{hill91} found a strong evolution of
  the RGs environment, because already at $z=0.5$ most powerful
  sources are located in rich clusters, unlike what is seen at lower
  redshifts.

Several radio surveys covering large areas and reaching the mJy flux
level became available in the last two decades, e.g., FIRST
\citep{becker95,helfand15} and NVSS \citep{condon98}. The studies of
the RGs environment based on these surveys broadly confirmed the
earlier results. \citet{best04} found that radio-AGN are
preferentially located in galaxy groups and poor-to-moderate richness
galaxy clusters. \citet{gendre13} provided further support to the
higher local galaxies density around FR~Is with respect to FR~IIs;
they also noted the possible presence of a link between the various
optical classes of RGs \citep{laing94}: high-excitation galaxies
(HEGs) are found almost exclusively in low-density environments while
low-excitation galaxies (LEGs) occupy a wider range of densities. By
using observations from the International Low Frequency Array (LOFAR;
\citealt{vanhaarlem13}) \citet{croston19} found a connection between
size and luminosity at 150 MHz of the brightest radio AGN with the
cluster richness. In contrast, \citet{massaro19}, hereafter
\citetalias{massaro19h}, concluded that regardless of their radio
morphological classification (FR~I or FR~II) and/or their optical
classification (LEGs of HEGs) RGs in the local universe live in
galaxy-rich large-scale environments that have similar characteristics
and richness. This different result is probably driven by the
different selection criteria of the samples.

These studies focused almost exclusively on the bright extended
RGs. However, the identification of the optical counterparts of
radio sources \citep{best05a,best12} in FIRST and NVSS showed that the
majority of them are associated with low redshift galaxies and are
unresolved \citep{baldi09}. This is a radical change in our view of
the radio sky, because earlier surveys (performed at lower frequency
and higher flux threshold) were dominated by sources extending over a
typical scale of hundreds of kpc (e.g., \citealt{hardcastle98}).  The
general lack of substantial extended radio emission suggested to define these
``compact'' sources as ``FR~0s'' \citep{ghisellini11,sadler14}, as a
convenient way to include them into the canonical \citet{fanaroff74}
classification scheme of radio galaxies (RGs).

The information on FR~0s is quite limited, even at the radio
frequencies used to classify them: the available radio data are of
poor resolution and with multi frequency data available only for the
FR~0s of higher flux density. As a consequence, it is still unclear
which is the nature of these compact sources and how they are related
to the other classes of RGs. 
In order to perform a systematic studies of FR~0s, \citet{baldi18h},
hereafter BCM18, selected a sample of 104 compact radio
sources, named \FRo, while \citet{capetti17} built a comparison sample
of extended, edge-darkened FR~I RGs (see Sect. 2 for further details
on the samples selection). The number density of \FRo\ sources is five
times higher than that of FR~Is, confirming quantitatively that they
represent the dominant population of radio sources in the local
Universe \citepalias{baldi18h}. \citeauthor{baldi18h} found that the
\FRo\ hosts are mostly luminous red early-type galaxies with large
black hole masses (10$^8 \lesssim$ M$_{\rm BH} \lesssim 10^9
$M$_\odot$). These properties are similar to those seen for the hosts
of FR~Is, they are just on average a factor 1.6 less massive but there
is a large overlap between the two mass distributions.

\citet{baldi15,baldi19} obtained high resolution multi-frequency radio
images of a sub sample of FR~0s extracted from \FRo. Although they
reach an angular resolution of 0\farcs3 (corresponding to a few
hundred pc), the majority of the FR~0s are still unresolved, while the
remaining extend only a few kpc. Most of them have flat spectra and
the ratio between the core and total emission in FR~0s is $\sim$ 30
times higher than that in FR~Is.

The comparison of optical line emission luminosity, a robust proxy of
the radiative power of the AGN, indicates that FR~0s share the same
range of FR~Is, but they have a median radio luminosity a factor
$\sim$30 smaller than that of the \FR\ \citepalias{baldi18h}. However,
there is no sharp boundary between the properties of FR~0s and FR~Is:
low-luminosity RGs form a continuous distribution, from the FR~0s at
the lowest ratios of radio/line luminosity, to the \FR\ sources at
intermediate ratios, and finally to the extreme values reached by the
most powerful FR~Is part of the Third Cambridge catalog
\citep{spinrad85}. If we instead consider only the emission from the
radio core, the FR~0s lie in the same region populated by the FR~Is,
indicating a common nature of the nuclei of the two groups of
sources. This similarity is also supported by available X-rays
observations \citep{torresi18}.

It therefore appears that, while the host galaxies and the nuclear
properties of FR~0s and FR~Is are very similar, their appearance in
the radio images is radically different. The origin of their different
nature still remains to be understood. For example, a scenario in
which FR~0s are young RGs that will all eventually evolve into
extended radio sources cannot be reconciled with the large space
density of FR~0s. FR~0s might instead be recurrent sources,
characterized by short phases of activity
\citepalias{baldi18h}. Finally, the jet properties of FR~0s might be
intrinsically different from those of the FR~Is, e.g., the former
class having lower bulk Lorentz factors \citep{baldi19}. The VLBI
observations obtained by \citet{cheng18} indicate a diversity of
relativistic beaming indicators among the sources of the sub-sample of
14 bright (with flux densities $>$ 50 mJy) FR~0s they analyzed.

In this work we extend our comparison of FR~0s and FR~Is by studying
their large-scale environment, probing distances up to 2 Mpc, testing
whether their different radio morphologies are related to, e.g., the
local densities of galaxies. The paper is organized as follows: in
Sect. 2 we describe the samples considered and whose environment is
studied in Sect. 3. The results are discussed in Sect. 4, while our
conclusions are given in Sect.  5.

Throughout the paper we assume the same cosmology as in M19, i.e.,
H$_0$ = 69.6 km s$^{-1}$ Mpc$^{-1}$, $\Omega_{\rm M}$ = 0.286, and
$\Omega_{\Lambda}$ = 0.714 \citep{bennett14}. Thus,
1\arcsec\ corresponds to 0.984 kpc at $z = 0.05$.

\begin{figure*}
\includegraphics[width=9.2cm]{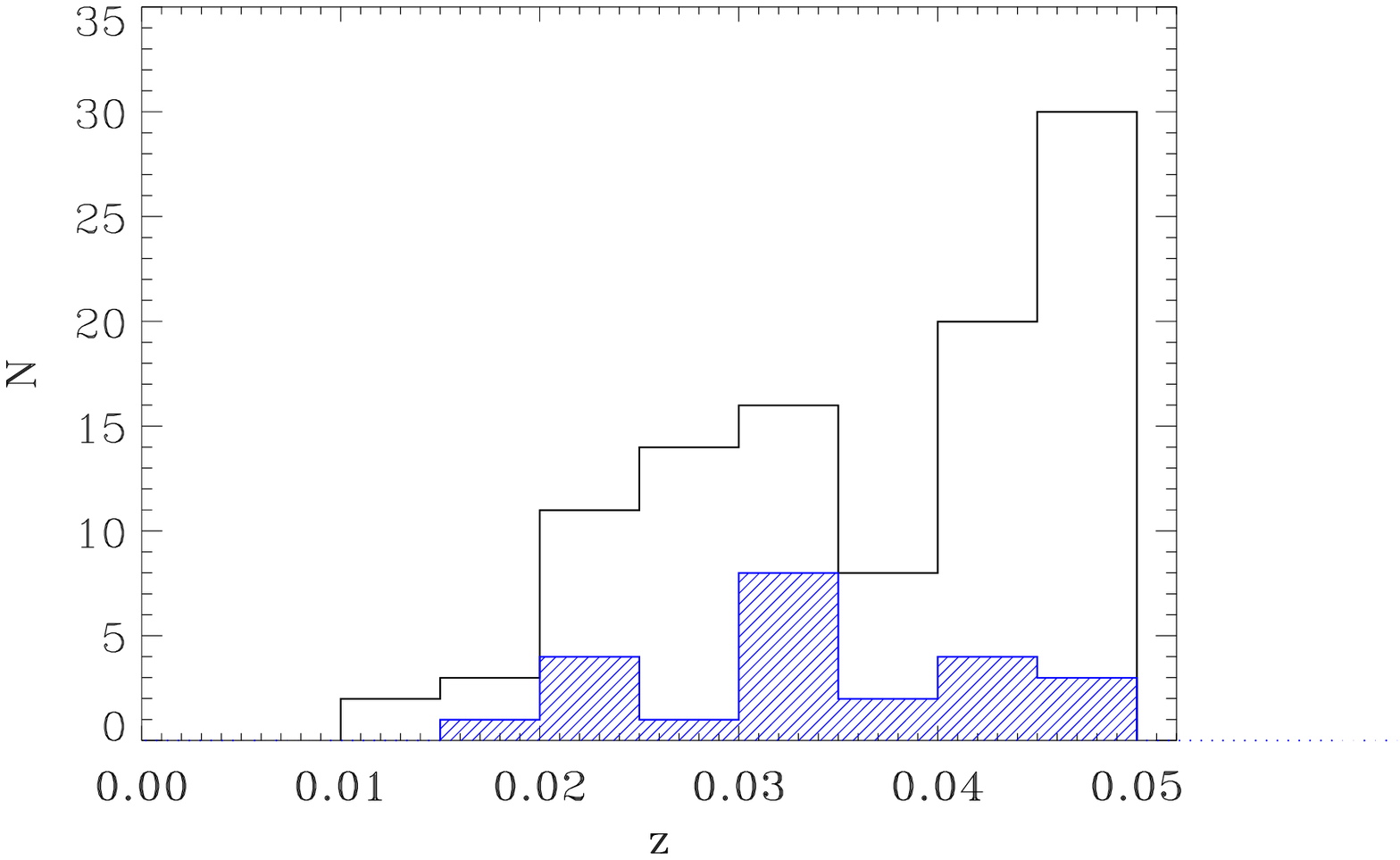}
\includegraphics[width=9.2cm]{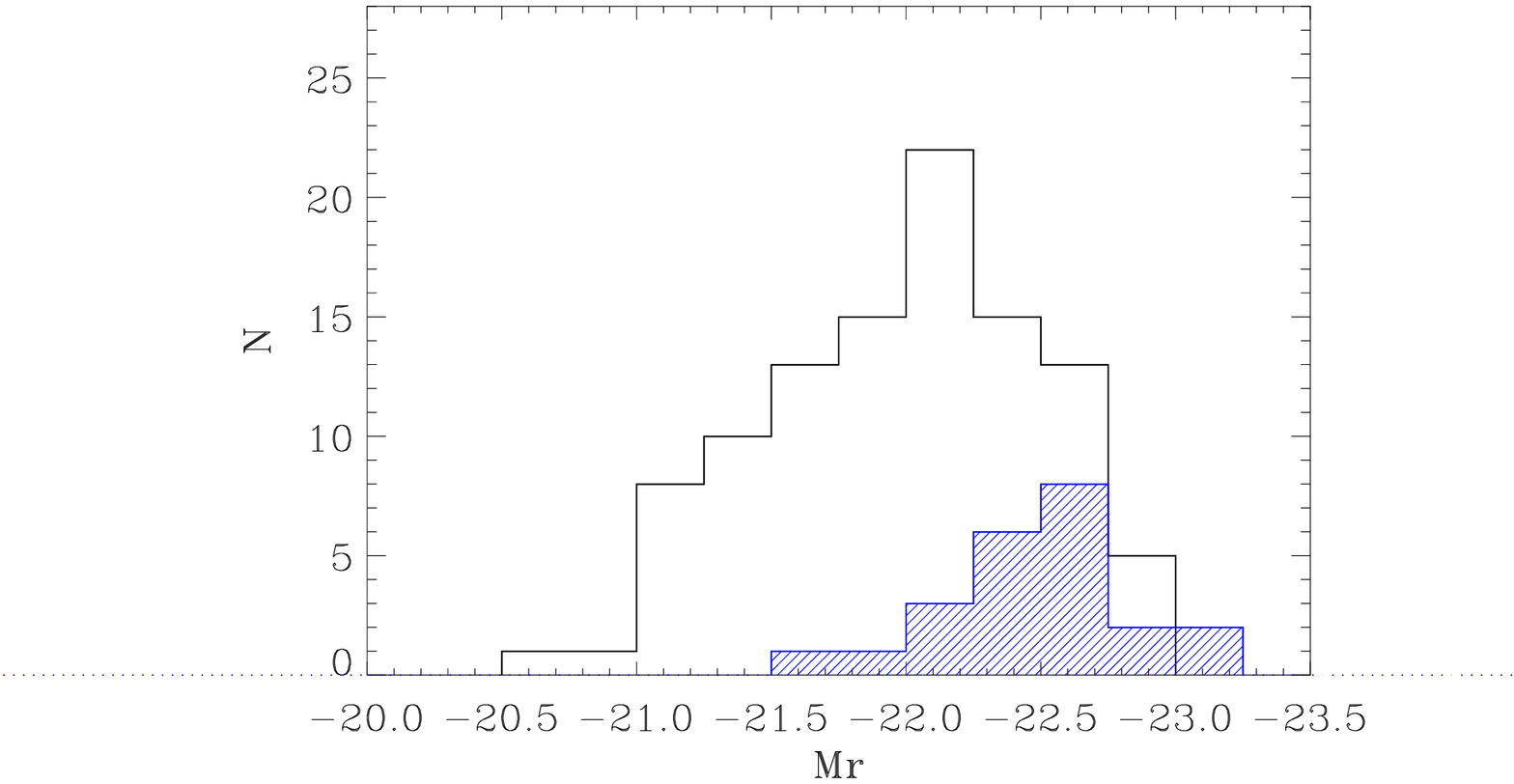}
\caption{Left: redshift distribution for the samples of RGs
  considered, black for the FR~0s, blue shaded for the FR~Is,
  including both \FR\ and \sFR\ objects.}
\label{hist}
\end{figure*}

\section{The samples}
The three samples of RGs we selected for the comparison of their
large-scale (up to 2 Mpc) environmental properties are those formed by
the compact \FRo\ sources with $z<0.05$, and the two catalogs of
edge-darkened sources, \FR, and \sFR, limiting to those with the same
redshift limit of the FR~0s.

We recently created a catalog of 219 FR~Is (\FR, \citealt{capetti17})
starting from the list of radio AGN produced by \citet{best12}. We
visually inspected the FIRST image for each individual source with
$z<0.15$ selecting those having radio emission with an edge darkened
morphology. We initially considered only those extending to a radius
$r$ larger than 30 kpc, i.e., well resolved sources at the 5$\arcsec$
resolution of the FIRST images within the redshift range of interest.
\citet{capetti17} also considered a second sample of 14 smaller FR~Is
(hereafter \sFR), having 10 $<$ r $<$ 30 kpc, but limiting to the
objects with $z <$ 0.05 to preserve a sufficient spatial
resolution. The final sample of low redshift FR~Is is composed by 23
objects, the 9 \FR\ with $z<0.05$ and the 14 \sFR.  From the point of
view of the optical spectroscopic classification, all \FR\ and
\sFR\ sources are low excitation galaxies (LEGs).

\FRo\ is instead composed by 104 compact RGs
\citepalias{baldi18h}.\footnote{The initial \FRo\ sample was formed by 108
  objects. The inspection of their NVSS images revealed the presence
  of low brightness diffuse emission, resolved out in the FIRST
  images, in four of them, that have then been removed from the
  catalog (see \citealt{baldi19} for further details).} They have been
selected from the \citeauthor{best12} catalog, imposing a redshift
limit of 0.05. We set a limit to the maximum deconvolved size of
4$\arcsec$, corresponding to a size $r \lesssim 2.5$ kpc. We also
required an optical spectroscopic classification as LEGs.

In the Appendix we provide three tables were we list the main
properties of the samples studied.

The redshift distributions of the samples considered are not
statistically distinguishable, with an average value of 0.037 for the
FR~0s and 0.036 for the FR~Is (see Fig. \ref{hist}, left panel, and
Table \ref{tab1}). Conversely, as already noted by \citetalias{baldi18h},
the FR~Is hosts are 0.66 magnitudes brighter than those of FR~0s
(Fig. \ref{hist}, right panel).

Following \citetalias{massaro19h}, in our analysis we also used a catalog of mock sources
(labeled as MOCK hereinafter) to estimate the efficiency of our
procedures. This has been created by shifting the positions of the
\FR\ sources by a random radius between 2$^\circ$ and 3$^\circ$ while
preserving their redshift distribution. The MOCK sample lists 4056
sources, 278 of which have $z<0.05$. The MOCK sample provides us with
a description of the environmental properties of random locations in
the local Universe to be compared with those derived from the sources
of our interest.

\section{Environmental properties}
\subsection{Estimates of the local galaxies density}

\citetalias{massaro19h} studied the environment of the sub sample 195
\FR\ sources lying in the central part of the SDSS footprint (see,
e.g., \citealt{ahn12}), the area covered by the catalog of groups and
clusters of galaxies produced by \citet{tempel12}, hereinafter T12,
which was used as reference for their analysis.

\citetalias{massaro19h} defined as ``cosmological neighbors'' all
galaxies lying within a region of given projected radius (they mostly
used a radius of 2 Mpc) and having a spectroscopic redshift $z$
differing by less than 0.005 from the radio galaxy in the center of
the field examined.  This choice corresponds to the maximum velocity
dispersion in groups and clusters of galaxies (see, e.g.,
\citealt{eke04}).  At the average redshift of the samples of RGs,
0.037, the spectroscopic limit of the SDSS (m$_r$=17.7) corresponds to
an absolute magnitude of M$_r$=-18.4. This luminosity is well below
the peak in the luminosity function of elliptical galaxies (e.g.,
\citealt{tempel11}) and $\sim$4 magnitudes below the median optical
luminosity of both FR~0s and FR~Is: at these low redshifts the SDSS
provides us with detailed information on environment.

\begin{figure}
\includegraphics[width=9.2cm]{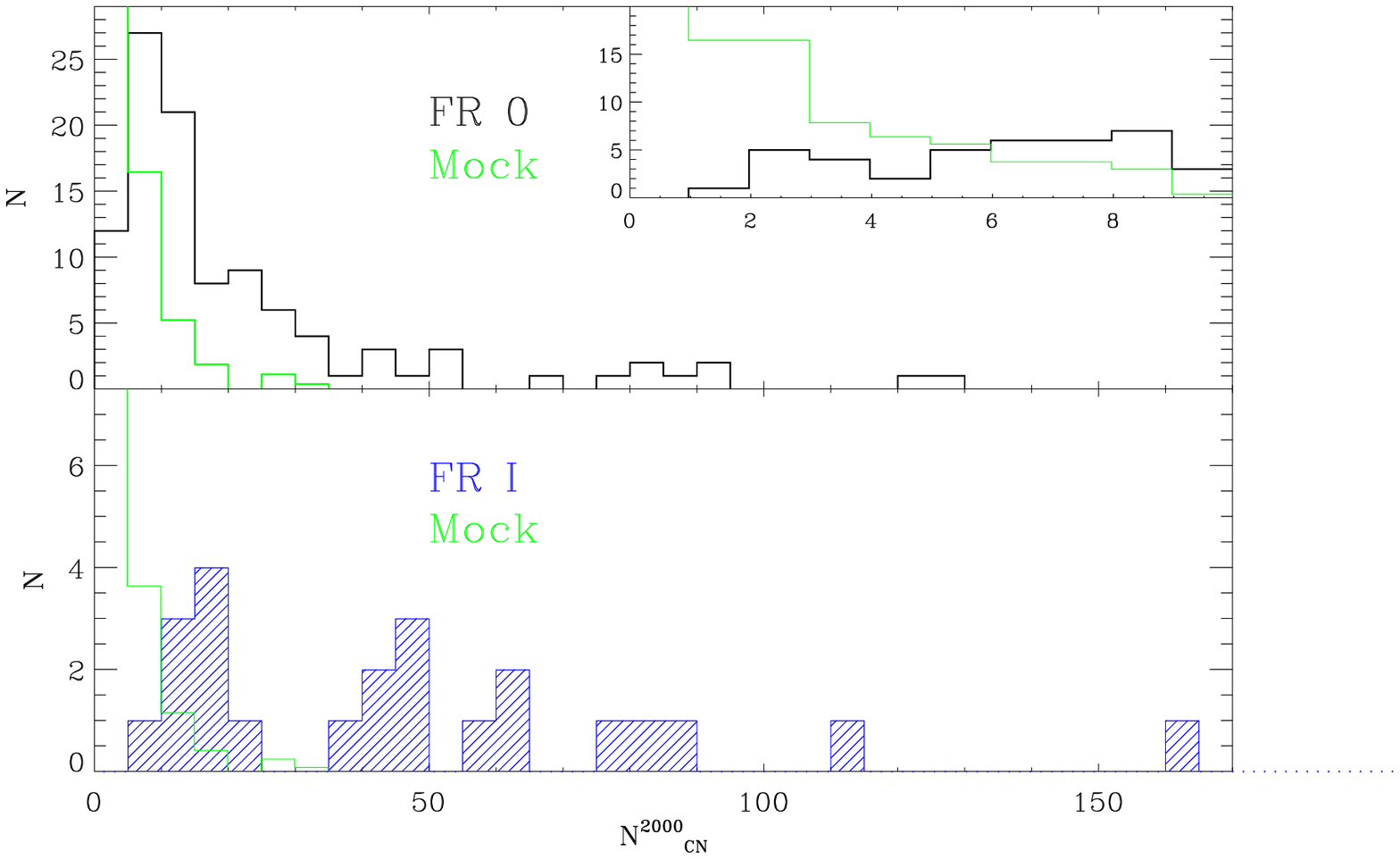}
\caption{Top: distribution of the number of cosmological neighbors for
  the FR~0s; the inset shows the same distribution with a smaller bin
  size, i.e., 1 instead of 5, to focus on the region of low \Ndue.  In
  green we show the \Ndue\ distribution for the sample of mock
  galaxies (see text for details) scaled by a factor 104/287, i.e.,
  normalized to the same area of the FR~0s histogram; the first bin
  contains 211 MOCKs. The inset is a zoom for low \Ndue\ for the both
  FR~0s and MOCKs; the first bin contains 85 MOCKs. Bottom: comparison
  of the \Ndue\ distributions for FR~Is (blue) and MOCKs (green).}
\label{f2000}
\end{figure}

The number of cosmological neighbors within 2 Mpc, \Ndue, spans a
large range, reaching values as high as 164, see Fig. \ref{f2000}. The
median values of \Ndue\ is 13 for the FR~0s and 44 for the FR~Is, and
the distributions of this parameter for the two classes differ
significantly, see Tab. \ref{tab1}. The same result is obtained when
using a radius of 1 Mpc, deriving \Nuno: the medians of this parameter
are 7 and 21 for FR~0s and FR~Is, respectively.

\begin{figure*}
\includegraphics[width=10.cm]{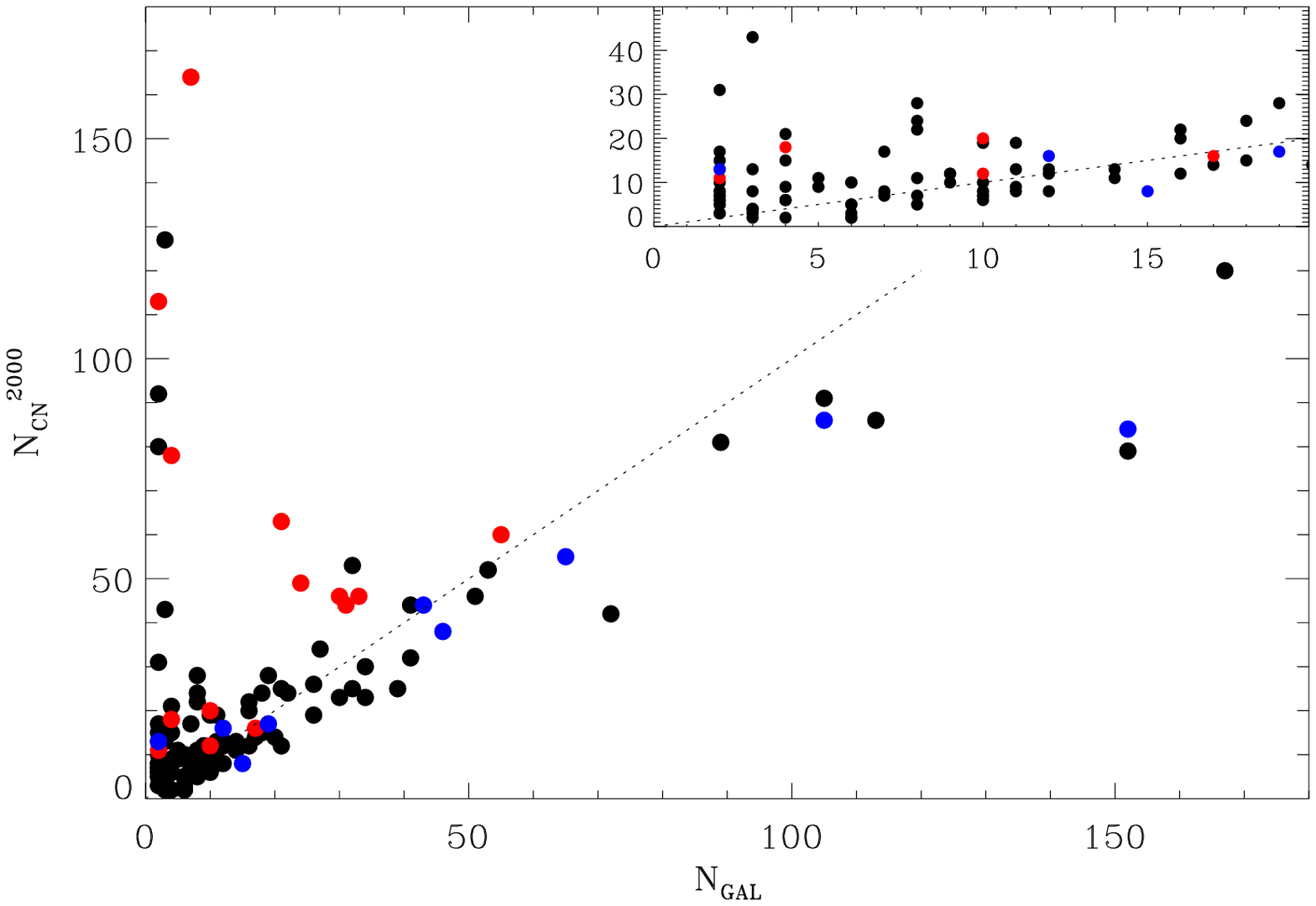}
\includegraphics[width=8.1cm,height=8.1cm]{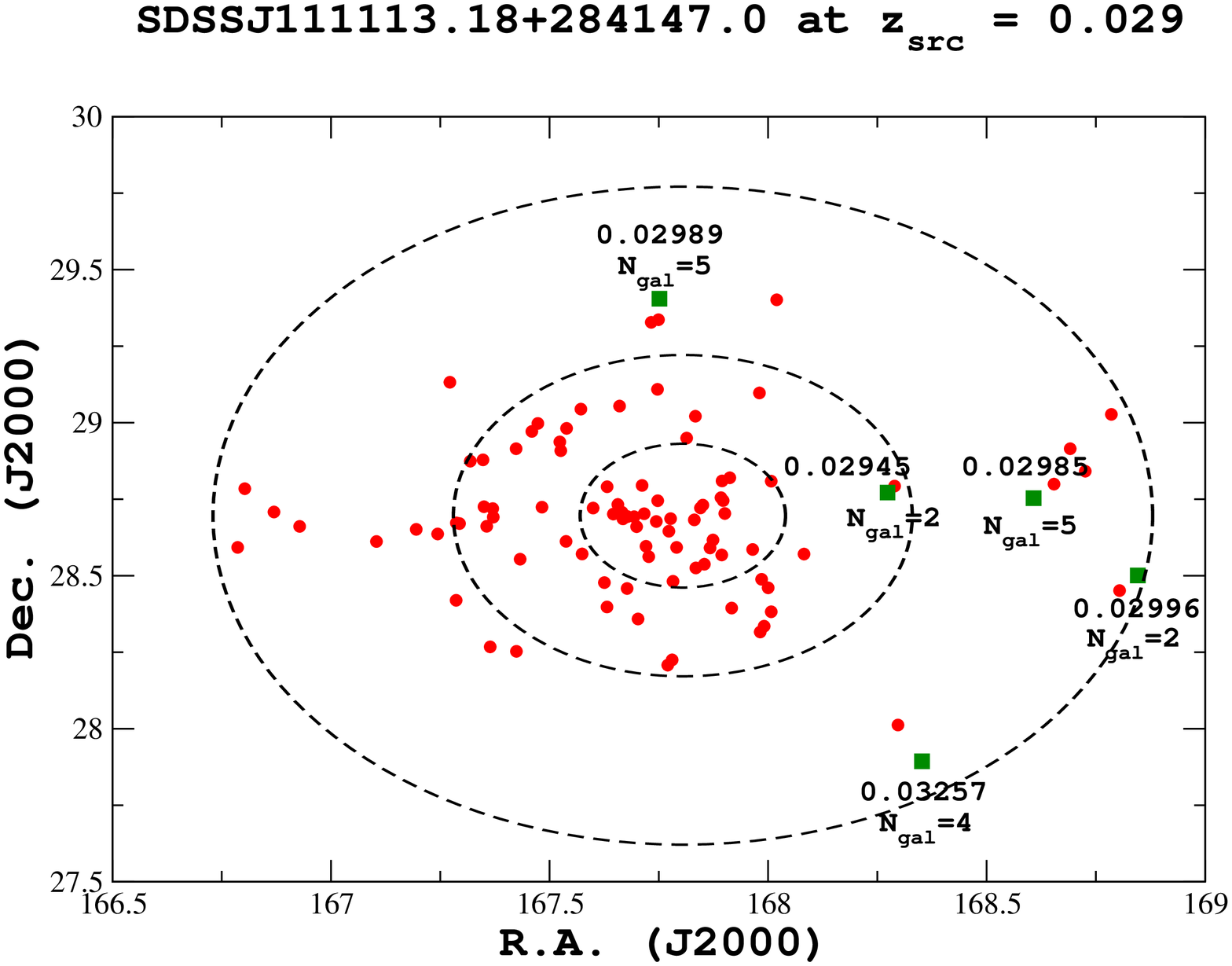}
\caption{Left: comparison between the number of cosmological neighbors
  within a diameter of 2000 kpc (\Ndue) and the number of galaxies
  ($N_{\rm gal}$) associated with the groups/clusters hosting the
  radio sources from T12. Black dots are FR~0s, blue and red are the
  \FR\ and \sFR\ objects, respectively. The inset in the top right
  corner is a zoom on the regions of low $N_{\rm gal}$ values. Right:
  the field around SDSS~J111113.18+284147.0 as example of the objects
  with a large discrepancy between $N_{\rm gal}$ and
  \Ndue. Cosmological neighbors are shown as red circles, while the
  green points mark the location of the closest (in projection) groups
  or clusters of galaxies with a redshift difference  $\Delta z <$0.005, listed
  in the T12 galaxy cluster/group catalog.}
\label{ngal-f2000}
\end{figure*}

We compare our results with those derived by cross-correlating our
lists of sources with published catalogs of galaxy clusters and
groups. More specifically we considered the T12 catalog created by
using a modified version of the Friends-of-Friends (FoF) algorithm
\citep{huchra82,tago10}.  In Fig. \ref{ngal-f2000} we compare the
number of cosmological neighbors, \Ndue, with the number of galaxies,
$N_{\rm gal}$ associated with the groups/clusters hosting the radio
sources according to T12 (this is possible for all the FR~0s but six,
as they are located outside the area covered by the T12
analysis). There is a general consistency between these two estimates
of the galaxies density. However, in several cases $N_{\rm gal}$ is
much smaller than \Ndue. For example, there are six objects with
$N_{\rm gal} < 5$ and \Ndue\ $> 60$. The field around one of them
(namely SDSS~J111113.18+284147.0) is shown in the right panel of
Fig. \ref{ngal-f2000}. T12 finds a group of two galaxies while we
found 92 cosmological neighbors, with a strong concentration of
sources $\sim 200$ kpc to the West. We found that by using the FoF
algorithm individual clusters of galaxies might be split into multiple
sub-structures which are not recognized as being part of a single
entity. When looking for the closest group/cluster at a given
position, the outcome could be an underestimate of the local galaxies
density. This effect is particularly severe at low redshift where
structures cover large areas of the sky. In this case, the counting of
cosmological neighbors is a more robust method for environmental
studies.

\begin{table*}
\caption{Summary of the statistical results}
\begin{tabular}{l| r r r r | r r r r | r }
\hline
      &       \multicolumn{4}{c}{Average} &  \multicolumn{4}{c}{Median} & \\
\hline
         &FR~0 & FR~I & sFR~I& FRICAT &FR~0 &FR~I&sFR~I&FRICAT &K.S. (P) \\
\hline
z       & 0.037&0.035& 0.034& 0.036&  0.039& 0.034& 0.034&0.032& \,\, (0.360) \\
M$_r$   &-21.97&-22.48&-22.39&-22.63&-22.05&-22.52&-22.39&-22.63&T (0.000) \\
\Ndue\      & 21.95&47.87& 52.86& 40.11&  13& 44& 46&38& T (0.002)  \\
\Nuno\      & 13.61&23.91& 21.71& 27.33&  7 & 21 & 22 & 17 &  T (0.002)\\ 
$N_{\rm gal}$    & 17.43&30.83& 17.86& 51.00&   8& 19& 17&43 &T (0.001) \\
log $\Sigma_{5}$& -4.74&-4.46&-4.50&-4.39&-4.72&-4.45&-4.45&-4.20 &T (0.026)\\
$N_{\rm cn,0.05}^{2000}$ & 15.68 &  23.74 & 24.07 & 23.22 & 8 & 16 & 16 & 15& T (0.016)  \\
\Rcn &   338& 347& 431& 216&  303& 398& 470 & 190 &  \,\, (0.186) \\
|\vpar|&   235& 192& 225& 141&  180& 193& 259 & 110 &  \,\, (0.770) \\
  \hline
\end{tabular}
\label{tab1}
\smallskip

\small{Column description: (1) parameter; (2 - 3) average value for
  the FR~0s, the FR~Is (split into (4 -5) \FR\ and \sFR\ sources); (6
  -7) median value for the FR~0s, the FR~Is (split into (8 -9)
  \FR\ and \sFR\ sources) (10) outcome of the Kolmogoroff-Smirnov test
  (and corresponding probability) T=the FR~0s and FR~Is populations
  differ significantly. The parameters considered are the redshift
  ($z$), the source absolute magnitude ($M_r$), the number of
  cosmological neighbors within 2 Mpc (\Ndue) and 1 Mpc (\Nuno), the
  number of galaxies ($N_{\rm gal}$) associated with the
  groups/clusters hosting the radio sources according to T12, the
  fifth nearest neighbor density ($\Sigma_5$), the number of
  cosmological neighbors within 2 Mpc ($N_{\rm cn,0.05}^{2000}$) when
  all sources are moved to a common redshift of 0.05, the projected
  distance in kpc (\Rcn) and the absolute value of the speed of light
  times the redshift difference in \kms\ (|\vpar|) from the average of
  the cosmological neighbors.}
\end{table*}

We analyzed the fields around the MOCKs with the same strategy used
for the RGs. In the bottom panel of Fig. \ref{f2000} we show as
green histogram the resulting distribution of \Ndue\ which shows a
strong concentration for low values of \Ndue. In particular, we find
that 95\% of the MOCKs correspond to \Ndue $<$ 11. This implies that a
value of \Ndue $>$ 11 has a probability of $<$5\% to occur by
chance. The strong difference in the distributions of \Ndue\ between
FR~Is, FR~0s, and the MOCKs (see the insets in the left panel of
Fig. \ref{f2000}) indicates that both classes of RGs are
located in regions of higher than average galaxies density.

\begin{figure}
\includegraphics[width=9.2cm]{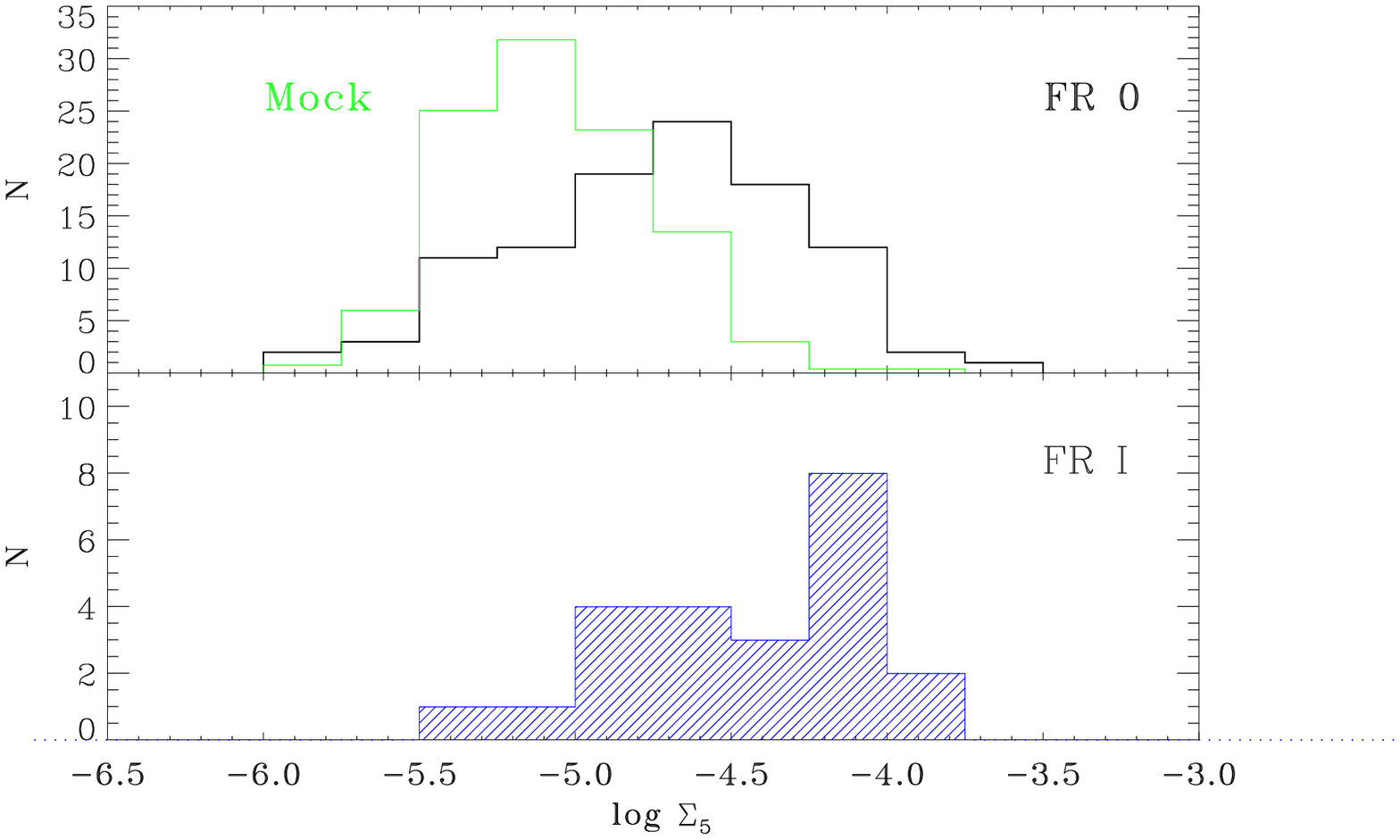}
\caption{Comparison of the $\Sigma_5$ parameter for (top panel) FR~0s
  (black) and MOCKs (green, scaled by a factor 104/278)), (bottom)
  distribution of $\Sigma_5$ for the FR~Is. }
\label{sigma5}
\end{figure}

We also estimated the projected galaxies density, following the
  approach of \citet{dressler80}, i.e, measuring the $\Sigma_k$ parameter.
  $\Sigma_k$ is defined as the ratio between the number of sources $k$
  and the projected area $\pi r_k^2$, where $r_k$ is the projected
  distance between the central galaxy and the $k$th nearest
  neighbor. More specifically, we estimated $\Sigma_5$, derived from
  the distance of the fifth closest candidate elliptical galaxy.
Candidate elliptical galaxies are optical sources lying within the 2
Mpc distance from the RG and having optical colors consistent with
those of quiescent ellipticals at the same redshift. The distributions
of $\Sigma_5$ for FR~0s and FR~Is differ significantly, with the
latter group showing a median value a factor $\sim$ 3 larger
(Fig. \ref{sigma5}). The distribution of $\Sigma_5$ for the MOCKs has
a median a factor $\sim 2$ lower than the average of the FR~0s. This
confirms that, overall, FR~0s are in located in an environment richer
than average.

All the estimators concur on the result that the local galaxies
density around FR~Is is a factor $\sim$ 2 - 3 larger than around
FR~0s. However, this ratio does not fully capture their different
environment. Fig. \ref{f2000} shows that the main characteristic of
the Mpc scale environment is the large fraction of FR~0s located in
poor groups of galaxies, an environment which FR~Is rarely
inhabit. More quantitatively, about 2/3 of the FR~0s (65/104) have
\Ndue $<$ 15, while this occurs for only 4 out 23 FR~Is
(17\%).

\begin{figure}
\includegraphics[width=9.2cm]{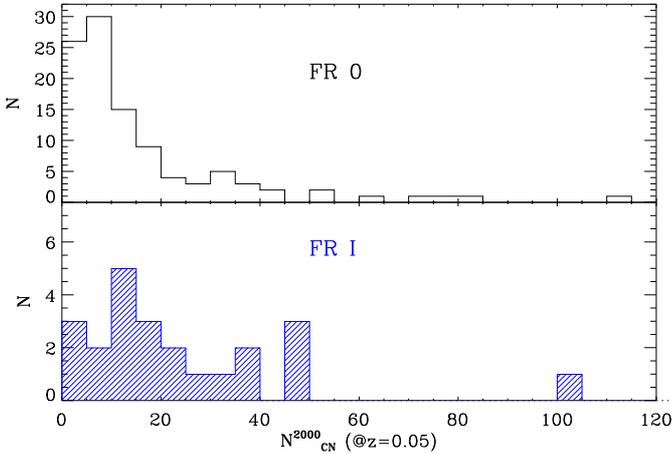}
\caption{Distribution of the simulated number of cosmological
  neighbors for the FR~0s and FR~Is (black and blue histograms,
  respectively) within 2 Mpc after moving all sources to a common
  redshift of 0.05.}
\label{f2000005}
\end{figure}

Despite the similarity in the distance distribution of FR~0s and FR~Is
we cannot exclude that some residual effect due to redshift is still
present and that this affects our results. To address this
possibility, we artificially `moved' all sources to a common redshift
of 0.05. The magnitudes of each source and of all its cosmological
neighbors are re-evaluated by considering the increased distance
modulus, with an average correction of $\sim 0.7$ magnitudes. All
neighbors which, after this flux dimming, fall below the threshold of
the SDSS spectroscopic selection (i.e., $r > 17.7$) are excluded from
the estimates of the local galaxies density. Effectively, this
strategy produces a list of cosmological friends with a fixed absolute
magnitude limit of $M_r < -19$, with the only drawback of reducing the
number of cosmological friends by an average factor 1.5.  In
Fig. \ref{f2000005} we compare the simulated number of cosmological
neighbors within 2 Mpc at a redshift of $z=0.05$, $N_{\rm
  cn,0.05}^{2000}$ of FR~0s and FR~Is: the two distributions still
differ at a high significance level (see Tab. \ref{tab1}).

We considered the possibility of a connection between the properties
of the environment, of the active nucleus, and of the host galaxy of
the RGs. In particular we tested the presence of a relation of
\Ndue\ with 1) the [O~III] line luminosity, 2) the host black hole
mass, and 3) the strength of the Dn(4000) index. The Spearman rank
test does not return any significant correlation. The only notable
result concerns the five FR~0s forming the tail of low black mass
values, $\log M_{\rm BH} < 7.8$, for which we find an even poorer
environment than the FR~0s population, with an average value \Nduemed
= 6.2.

\begin{figure*}
\includegraphics[width=9.2cm]{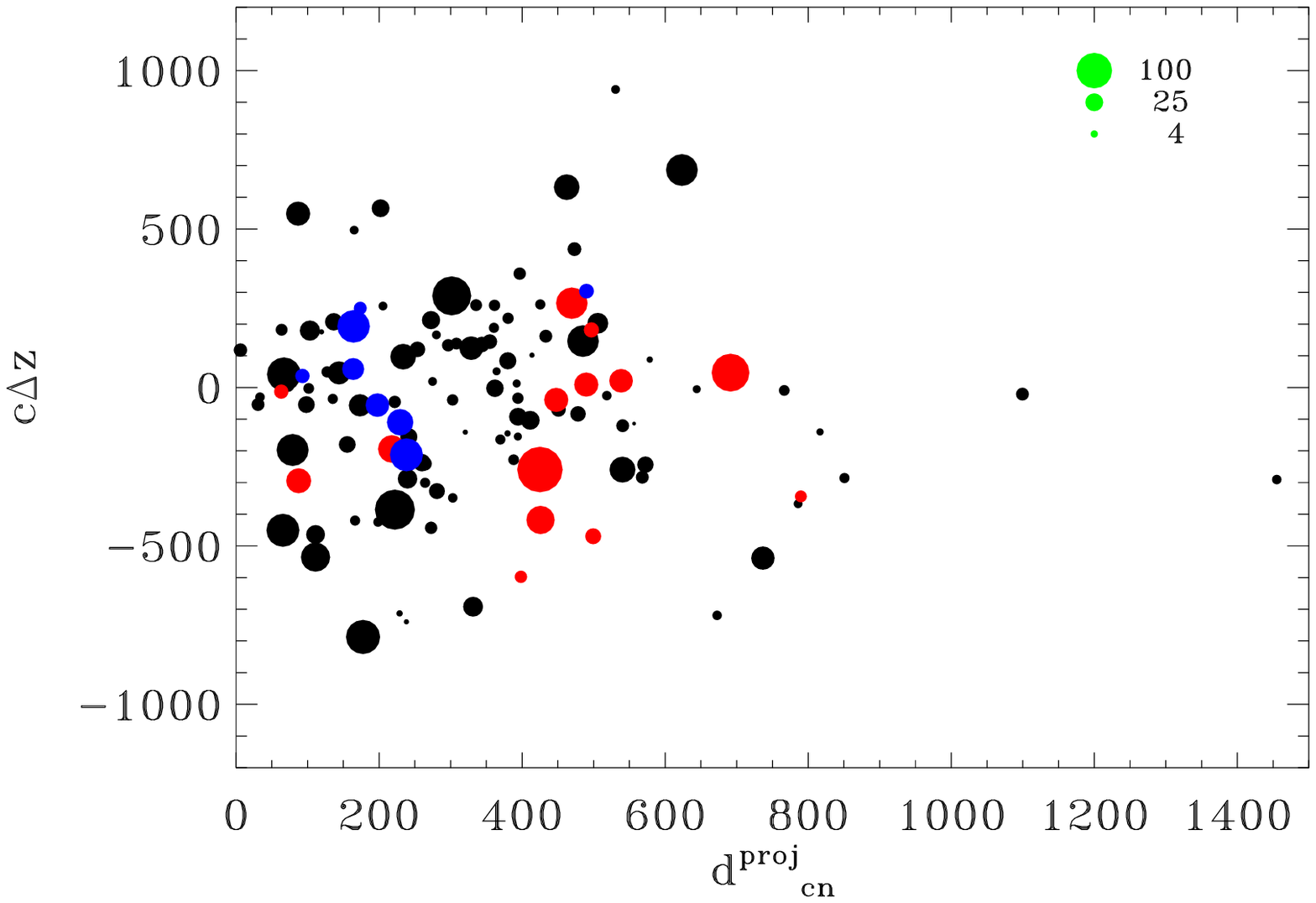}
\includegraphics[width=9.2cm]{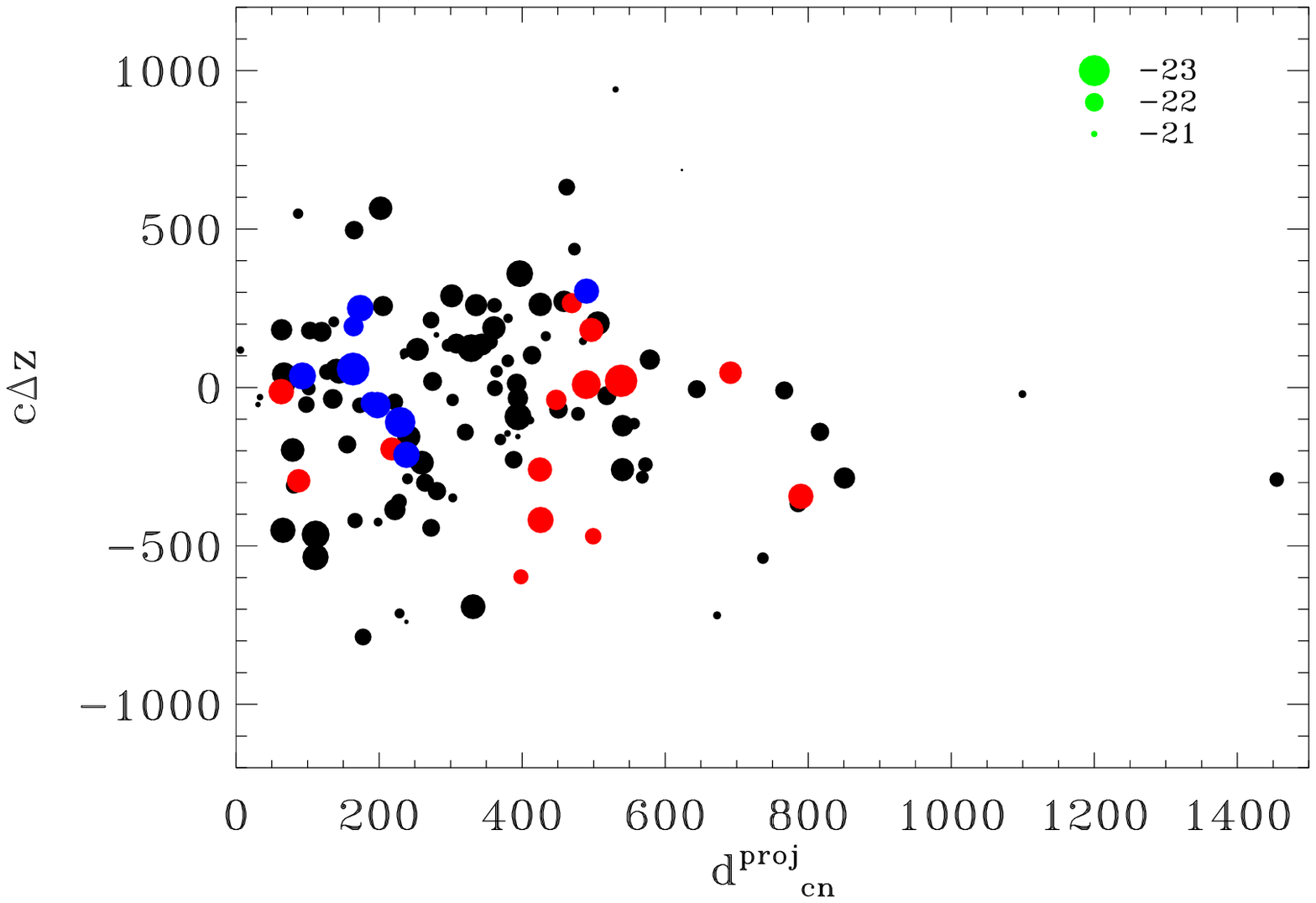}
\caption{Projected distance (in kpc) versus the redshift difference
  times the speed of light (in \kms) for each RG from the average of
  its cosmological neighbors. In the left panel the symbol size is
  proportional to \Ndue\ (see the coding on the top right) while in
  the right panel is proportional to the host absolute magnitude.}
\label{rdv}
\end{figure*}

\subsection{Location of the RGs within the group/cluster of
  galaxies}

Using optical observations, the local density of galaxies is not the
only parameter which defines the environmental properties as it is
also important to establish in which location within a, e.g., cluster
of galaxies a given source is located. The location of an AGN within
the group/cluster of galaxies might have a profound effect on the
level of accretion and, consequently, on its nuclear power (see, e.g.,
\citealt{koulouridis18}). For example, by exploring the properties of
early-type galaxies in the Virgo cluster, \citet{vattakunnel10} found
a suggestive trend between jet power and location within the cluster.
A similar result was found by \citet{croston19}.

We then estimated the projected distance, \Rcn, of each RG from the
average of the positions in the sky of all its cosmological neighbors
within 2 Mpc. Similarly, we estimated the difference between their
redshift with respect to the average of the cosmological neighbors,
\vpar. These two quantities are reported in Fig. \ref{rdv} and a
statistical summary is given in Tab. 1. We found \Rcn\ $<$900 kpc
  and |\vpar| $<$ 800 \kms for all but three of the RGs
  considered.

Both the average distance ($\sim$340 kpc) and the difference in
velocity ($\sim 200$ \kms) are similar for FR~0s and FR~Is and the
distributions for the two classes are not statistically
distinguishable. The same result is obtained when considering the
distribution of cosmological neighbors within 1 Mpc, with \Rcn = 243
kpc and 250 kpc for the FR~0s and FR~Is, respectively.  There is a
difference between the average values between \FR\ and \sFR\ sources,
both indicating that the former are closer to the center of the
group/cluster, but the small number statistics prevents to draw a firm
conclusion. Actually, when restricting to the cosmological neighbors
within 1 Mpc this difference disappears.

We also explored the possibility that the location of the radio
sources depends on its optical luminosity or on the richness of the
galaxies structure. For this reason the symbol sizes in the two panels
of Fig. \ref{rdv} are proportional to \Ndue\ and $M_r$, respectively:
we do not find any apparent dependency between these quantities. More
quantitatively, we experimented whether a cut-off at low values of
\Ndue\ and/or $M_r$ affects the results, but this not the case: in
particular the median values of \Rcn\ and \vpar\ do not change
significantly.

We conclude that there is often a large displacement (of the order of
200-300 kpc) of the RGs hosts from the average of the projected location of the
cosmological neighbors and that FR~0s and FR~Is do not show
significant differences in this respect.

On the other hand the host galaxies of the RGs are invariably the most
luminous galaxy among the cosmological neighbors within 2 Mpc. This
might an indication of a substantial complexity in the distribution of
galaxies, possibly indicative of the the presence of sub-structures,
not yet completely relaxed. Alternatively, the average of the galaxy
locations might not be accurately tracing the location of the center
of the galaxies structure.

\section{Discussion}
The main conclusion on the environmental properties of the three
samples of low redshift RGs is that FR~0s live in regions of lower
galaxies density with respect to FR~Is, independently on the method
used. This is driven by the small fraction of FR~Is located in groups
formed by less than 15 galaxies, an
environment which, conversely, is typical of FR~0s. The poorer
environment of FR~0s with respect to the FR~Is is the first
significant difference between these two classes of RGs, leaving aside
the defining characteristic of FR~0s, i.e., the lack of substantial extended
radio emission.

One possibility to account for the connection between environment and
properties of the extended radio emission is related to the adiabatic
losses of the radio emitting plasma (e.g., \citealt{longair94}). In a
poorer environment density and pressure of the external medium are
reduced with respect to regions of higher galaxies density: this
causes a faster lateral expansion of the jets, stronger adiabatic
losses and, consequently, a reduced emissivity. However, the high
resolution radio observations by \citet{baldi19} show that most FR~0s
do not reach sizes of even $\sim$ 1 kpc: at these small scales the
external gas in which they expand is still well within the core of hot
corona of their host. The similarity of the host galaxies of FR~0s and
FR~Is suggests that also their coronae will have similar properties,
based on the connection between optical and X-ray luminosity
\citep{fabbiano92}. A large spread in the X-ray properties for
galaxies of similar absolute magnitude exists, but it does not appear
to be closely connected with the local galaxies density \citep{su15}.
The possibility that the separation between FR~0s and FR~Is is driven
by differences in their hosts hot gas content appears contrived: the
paucity of extended radio emission in FR~0s is more likely to be an
intrinsic property of these sources.  Nonetheless, it would be
important to test this conclusion with X-ray imaging of these low
redshift RGs.

\citet{baldi15,baldi19} suggested that FR~0s are associated with jets
of lower bulk Lorentz factor $\Gamma$ with respect to FR~Is, thus
reducing their ability to penetrate the ambient
medium. \citet{baldi15} proposed that a high $\Gamma$ jet, leading to
a FR~I morphology, is only produced when the BH spin is close to its
maximum value, following the suggestions of a dependence between the
BH spin and $\Gamma$
\citep{McKinney05,Tchekhovskoy10,Chai12,Maraschi12}. FR~0s could be
associated with BHs of lower spin.

The origin of the connection between BH spin and environment can have
two explanations depending on whether the spin evolution is mainly
driven by accretion or black hole mergers. 

Within the first option, \citet{garofalo19} included the FR~0s into an
evolutionary framework for RGs: FR~0s represent the class of sources
formed during the transition from highly retrograde spinning (with
respect to the accretion disk) BHs associated with the powerful
FR~IIs, to the prograde BHs of the less powerful FR~II LEG or
FR~I. \citeauthor{garofalo19} ascribe the change of BH spin to the
angular momentum of the accreting material: a low spin FR~0 evolves
into a highly spinning source, i.e., a FR~I, when the accreting
material reaches $\sim$30\% of the initial BH mass. We note that RGs
do not need to follow this full evolution, starting as powerful
FR~IIs: FR~0s might form associated with low spin black holes and then
evolve into FR~Is. The connection between environment and the classes
of RGs requires, in this scheme, a positive link between local
galaxies density and accretion rate. If this is the case, the low BH
spin phase would last longer in poorer environment than in regions of
higher galaxies density. This would generate a connection between the
environment and the jet speed and power. This connection can be seen
only on a statistical basis, not on individual objects. In fact, FR~0s
would inhabit preferentially a poor environment, but can also be found
in clusters of galaxies if they formed recently and do not have yet
accreted a sufficient amount of mass onto their central BH to turn
into an FR~I. Similarly, FR~Is might be seen also in poor groups of
galaxies. There is evidence that the accretion rate in RGs is
controlled by the amount of hot gas available in the circumnuclear
regions \citep{allen06,balmaverde08}, but the process that would
eventually connect the scale at which this is measured (typically
$\lesssim$ 100 pc) and to the Mpc environment remains to be
understood. In this respect, it would be very important to be able to
assess the location of RGs with respect the center of the
gravitational well.

Alternatively, the BHs spin distribution is set mainly by their
mergers history. The simulations performed by \citet{dubois14}
indicate that indeed the most massive BHs (M$_{\rm BH} \gtrsim 10^8
$M$_\odot$), in particular those associated with gas poor galaxies
(such as the low redshift RGs we are considering), acquire most of
their mass through BH coalescence. An analysis of how the BH spin
distribution is related to environment, focusing on the massive
early-type galaxies, is needed in order to link the observed
differences between FR~0s and FR~Is to the results of numerical
simulations.

\section{Summary and conclusions}
We compared the environment on Mpc scale of FR~0s and FR~Is taking
advantage of the information of the local galaxies distribution
provided by the SDSS. The samples considered are formed by 104 objects
from \FRo, 14 from \sFR, and 9 from \FR\ (limiting to those with
$z<0.05$, the same redshift limit of the other two samples), the
latter two catalogs formed by FR~Is of different linear size. The
redshift distributions of samples considered do not differ
significantly and this enables us to perform a direct comparison
between their environment free from biases due to distance.

Following the methods described in \citetalias{massaro19h}, we used as
tracer of the local density of galaxies the number of cosmological
neighbors, i.e., the galaxies located within a given radius (usually 2
Mpc, but our results are unchanged when using smaller radii, e.g., 1
Mpc) and whose spectroscopic redshift differ by less than 0.005 from
the sources of our interest. The median number of cosmological
neighbors is a factor $\sim$ 3 larger for FR~Is than FR~0s. The same
conclusion is reached when considering other estimators, the fifth
nearest neighbor density $\Sigma_5$ or the number of galaxies
associated with the RGs according to a catalog of galaxy clusters and
groups. This difference is due to the large fraction (63\%) of FR~0s
located in groups of galaxies formed by less than 15 sources, where we
only find 17\% of the FR~Is. The poorer environment of FR~0s with
respect to the FR~Is is the first significant difference found between
besides the defining property of FR~0s, i.e., the lack of substantial  extended
radio emission.

The possibility that this link is due stronger adiabatic losses, that
might cause a lower jet brightness in FR~0s, appears to be contrived:
their jets are confined within the hot corona of their hosts and are
therefore unaware of the distribution of the external medium on larger
scales.

Our results suggest a connection between environment and jet power,
driven by a common link with the BH spin. There are two possibilities:
low spin RGs (i.e., the FR 0s) might evolve into high spin FR Is due
to accretion: in an environment of lower density the FR 0s phase would
last longer. Alternatively, the BH spin distribution results from
galaxies mergers and the BH coalescence: the most massive BHs located
in gas poor galaxies indeed acquire most of their mass through
coalescence. The role of environment on the BH spin evolution remains
to be fully investigated.

Our analysis of the environment is only based on optical data.
Clearly, X-rays observations of these low redshift RGs are crucial for
an independent and complementary analysis of their environment. In
fact, the X-rays luminosity and temperature of the inter-galactic
medium (IGM) provide an estimate of the total mass of galaxies
structures. Furthermore, we found that while both FR~0s and FR~Is are
always associated with the brightest galaxy among the cosmological
neighbors, they do not appear to be located at the barycenter of their
neighbors. This might an indication that the distribution of galaxies
is not yet fully relaxed or that the optical average might not be
accurately tracing the location of the center of the galaxies
structure. The IGM distribution derived from X-rays images might
provide a clearer answer to this problem. Finally, such data would
also enable us to compare the properties of the RGs hot coronae in
which their jets propagate.

\begin{acknowledgements}
We thank M. Volonteri, A. Paggi, I. Pillitteri, and
R. Campana, C. C. Cheung and A. Tramacere for useful comments and discussion.

This work is supported by the ``Departments of Excellence 2018 -
2022'' Grant awarded by the Italian Ministry of Education, University
and Research (MIUR) (L. 232/2016). This research has made use of
resources provided by the Compagnia di San Paolo for the grant awarded
on the BLENV project (S1618\_L1\_MASF\_01) and by the Ministry of
Education, Universities and Research for the grant
MASF\_FFABR\_17\_01. This investigation is supported by the National
Aeronautics and Space Administration (NASA) grants GO4-15096X,
AR6-17012X and GO6-17081X. F.M. acknowledges financial contribution
from the agreement ASI-INAF n.2017-14-H.0.  Funding for SDSS and
SDSS-II has been provided by the Alfred P. Sloan Foundation, the
Participating Institutions, the National Science Foundation, the
U.S. Department of Energy, the National Aeronautics and Space
Administration, the Japanese Monbukagakusho, the Max Planck Society,
and the Higher Education Funding Council for England. The SDSS Web
Site is http://www.sdss.org/. The SDSS is managed by the Astrophysical
Research Consortium for the Participating Institutions. The
Participating Institutions are the American Museum of Natural History,
Astrophysical Institute Potsdam, University of Basel, University of
Cam- bridge, Case Western Reserve University, University of Chicago,
Drexel University, Fermilab, the Institute for Advanced Study, the
Japan Participation Group, Johns Hopkins University, the Joint
Institute for Nuclear Astrophysics, the Kavli Institute for Particle
Astrophysics and Cosmology, the Korean Scientist Group, the Chinese
Academy of Sciences (LAMOST), Los Alamos National Laboratory, the Max-
Planck-Institute for Astronomy (MPIA), the Max-Planck- Institute for
Astrophysics (MPA), New Mexico State University, Ohio State
University, University of Pittsburgh, University of Portsmouth,
Princeton University, the United States Naval Observatory, and the
University of Washington. 
\end{acknowledgements}

\appendix
\section{Appendix}

\onecolumn
\begin{center}
\begin{longtable}{l r r r r r r r r r r}

\caption[Properties of the \FRo\ sample.]{Properties of the \FRo\ sample.} 
\label{tab} \\

\hline \hline 
name &  z & M$_{\rm r}$ & \Ndue & $N_{\rm cn,0.05}^{2000}$ & \Nuno & N$_{\rm gal}$ & $\log \Sigma_5$ & \Rcn & \vpar \\
\hline	
\endfirsthead

\multicolumn{3}{c}{{\tablename} \thetable{} -- Continued} \\[0.5ex]
\hline \hline 
name &  z & M$_{\rm r}$ & \Ndue &  $N_{\rm cn,0.05}^{2000}$& \Nuno & N$_{\rm gal}$ & $\log \Sigma_5$ & \Rcn & \vpar \\
\hline
\endhead

\hline
  \multicolumn{10}{c}{{Continued on Next Page}} \\
\endfoot

  \\[-1.8ex] 
\endlastfoot

SDSS~J010852.48-003919.4  & 0.045 & -21.42 &   7 &  7 &  6 & --- &  -5.34 &  303 & -348\\
SDSS~J011204.61-001442.4  & 0.044 & -21.62 &   1 &  0 &  1 & --- &  -5.91 &  556 & -114\\
SDSS~J011515.78+001248.4  & 0.045 & -21.96 &  39 & 30 & 39 & --- &  -3.71 &  173 &  -56\\
SDSS~J015127.10-083019.3  & 0.018 & -21.13 &  13 &  7 & 13 & --- &  -4.04 &   30 &  -53\\
SDSS~J020835.81-083754.8  & 0.034 & -22.19 &   2 &  2 &  2 & --- &  -4.59 &  413 &  101\\
SDSS~J075354.98+130916.5  & 0.048 & -22.31 &  11 & 11 &  8 &  14 &  -5.02 &  308 &  138\\
SDSS~J080716.58+145703.3  & 0.029 & -21.82 &   9 &  8 &  8 &  11 &  -4.49 &  101 &   -3\\
SDSS~J083158.49+562052.3  & 0.045 & -22.00 &  17 & 14 &  2 &   2 &  -4.97 &  354 &  144\\
SDSS~J083511.98+051829.2  & 0.046 & -22.06 &   6 &  6 &  4 &   2 &  -5.37 &  785 & -366\\
SDSS~J084102.73+595610.5  & 0.038 & -22.13 &  12 &  8 &  7 &   9 &  -4.35 &  272 & -443\\
SDSS~J084701.88+100106.6  & 0.048 & -22.15 &   5 &  5 &  2 &   2 &  -5.65 &  644 &   -5\\
SDSS~J090652.79+412429.7  & 0.027 & -21.56 &  24 &  9 & 21 &  18 &  -4.20 &  136 &  207\\
SDSS~J090734.91+325722.9  & 0.049 & -21.73 &  13 & 13 &  3 &   3 &  -5.58 &  568 & -283\\
SDSS~J090937.44+192808.2  & 0.028 & -21.58 &  30 & 20 & 23 &  34 &  -4.71 &  239 & -288\\
SDSS~J091039.92+184147.6  & 0.028 & -22.15 &   8 &  2 &  8 &  10 &  -4.68 &  264 & -300\\
SDSS~J091601.78+173523.3  & 0.029 & -22.44 &  42 & 26 & 38 &  72 &  -4.21 &  143 &   46\\
SDSS~J091754.25+133145.5  & 0.050 & -21.14 &   6 &  5 &  3 &   4 &  -5.14 &  280 &  166\\
SDSS~J093003.56+341325.3  & 0.042 & -22.00 &   5 &  4 &  5 &   8 &  -5.03 &   80 &   29\\
SDSS~J093346.08+100909.0  & 0.011 & -21.30 &  14 &  2 & 14 &  17 &  -5.15 &    6 &  118\\
SDSS~J093938.62+385358.6  & 0.046 & -21.70 &  10 & 10 &  2 &   6 &  -4.75 &  302 &  -39\\
SDSS~J094319.15+361452.1  & 0.022 & -21.79 &  12 &  5 & 12 &  21 &  -4.32 &   72 &    5\\
SDSS~J100549.83+003800.0  & 0.021 & -21.21 &   7 &  3 &  7 &  10 &  -4.31 &   33 &  -30\\
SDSS~J101329.65+075415.6  & 0.046 & -22.23 &   7 &  6 &  4 &   7 &  -5.11 &  518 &  -25\\
SDSS~J101806.67+000559.7  & 0.048 & -21.69 &   5 &  4 &  3 &   2 &  -5.13 &  364 &   51\\
SDSS~J102403.28+420629.8  & 0.044 & -21.81 &  19 & 16 &  9 &  10 &  -4.45 &  478 &  -83\\
SDSS~J102511.50+171519.9  & 0.045 & -22.62 &  34 & 31 & 14 &  27 &  -4.51 &  505 &  202\\
SDSS~J102544.22+102230.4  & 0.046 & -22.15 &  22 & 19 &  5 &   8 &  -4.79 &  155 & -179\\
SDSS~J103719.33+433515.3  & 0.025 & -21.98 &   3 &  1 &  2 &   3 &  -4.38 &   80 & -309\\
SDSS~J103952.47+205049.3  & 0.046 & -22.18 &   4 &  4 &  2 &   3 &  -5.13 &  816 & -140\\
SDSS~J104028.37+091057.1  & 0.019 & -22.07 &   2 &  0 &  2 &   4 &  -4.31 &  320 & -140\\
SDSS~J104403.68+435412.0  & 0.025 & -21.77 &  15 &  8 & 12 &  18 &  -4.25 &  264 & -240\\
SDSS~J104811.90+045954.8  & 0.034 & -22.29 &   5 &  4 &  5 &   6 &  -4.67 &  392 &   12\\
SDSS~J104852.92+480314.8  & 0.041 & -22.12 &   9 &  7 &  1 &   5 &  -4.95 &  388 & -227\\
SDSS~J105731.16+405646.1  & 0.025 & -22.29 &   8 &  5 &  7 &  11 &  -4.41 &  135 &  -36\\
SDSS~J111113.18+284147.0  & 0.029 & -22.05 &  92 & 54 & 71 &   2 &  -3.93 &  177 & -787\\
SDSS~J111622.70+291508.2  & 0.045 & -22.74 &  86 & 75 & 50 & 113 &  -4.43 &   65 & -450\\
SDSS~J111700.10+323550.9  & 0.035 & -22.04 &  26 & 18 & 15 &  26 &  -4.31 &  272 &  212\\
SDSS~J112029.23+040742.1  & 0.050 & -22.47 &  19 & 18 &  6 &  11 &  -5.38 &  343 &  135\\
SDSS~J112256.47+340641.3  & 0.043 & -22.93 &  44 & 30 & 21 &  41 &  -3.84 &  328 &  124\\
SDSS~J112625.19+520503.5  & 0.048 & -21.32 &  13 & 11 &  2 &   2 &  -5.40 & 1099 &  -20\\
SDSS~J112727.52+400409.4  & 0.035 & -21.20 &   6 &  3 &  3 &   2 &  -4.95 &  530 &  940\\
SDSS~J113449.29+490439.4  & 0.033 & -22.63 &  91 & 53 & 63 & 105 &  -4.25 &   67 &   41\\
SDSS~J113637.14+510008.5  & 0.050 & -21.93 &   8 &  8 &  2 &   2 &  -5.58 &  166 & -419\\
SDSS~J114230.94-021505.3  & 0.047 & -22.22 &   6 &  5 &  3 &   4 &  -5.22 &  274 &   19\\
SDSS~J114232.84+262919.9  & 0.030 & -22.58 &  23 & 13 & 16 &  30 &  -4.42 &  241 & -155\\
SDSS~J114804.60+372638.0  & 0.042 & -22.54 &  19 & 14 & 10 &  26 &  -4.95 &  253 &  120\\
SDSS~J115531.39+545200.4  & 0.050 & -21.88 &  21 & 21 &  8 &   4 &  -4.85 &  572 & -243\\
SDSS~J120551.46+203119.0  & 0.024 & -21.34 &  79 & 39 & 58 & 152 &  -4.18 &  485 &  146\\
SDSS~J120607.81+400902.6  & 0.037 & -22.44 &  13 & 10 &  6 &  11 &  -5.06 &  540 & -120\\
SDSS~J121329.27+504429.4  & 0.031 & -22.85 &  12 &  4 &  8 &   9 &  -4.06 &  396 &  359\\
SDSS~J121951.65+282521.3  & 0.027 & -21.12 &  52 & 30 & 37 &  53 &  -4.27 &  233 &   97\\
SDSS~J122421.31+600641.2  & 0.044 & -22.42 &  11 &  8 &  6 &   5 &  -5.38 &   63 &  182\\
SDSS~J123011.85+470022.7  & 0.039 & -22.62 &  24 & 19 & 17 &  22 &  -4.13 &  259 & -237\\
SDSS~J124318.73+033300.6  & 0.048 & -22.35 &  10 &  9 &  6 &   9 &  -5.37 &  394 &  -33\\
SDSS~J124633.75+115347.8  & 0.047 & -22.59 &   8 &  8 &  6 &  12 &  -4.23 &  425 &  262\\
SDSS~J125027.42+001345.6  & 0.047 & -21.23 &   3 &  3 &  3 &   2 &  -5.40 &  379 & -145\\
SDSS~J125409.12-011527.1  & 0.047 & -21.87 &   7 &  5 &  1 &   2 &  -5.36 & 1455 & -290\\
SDSS~J130404.99+075428.4  & 0.046 & -22.94 &  28 & 27 & 17 &  19 &  -4.75 &  111 & -464\\
SDSS~J130837.91+434415.1  & 0.036 & -22.57 &  53 & 40 & 18 &  32 &  -4.87 &  540 & -259\\
SDSS~J133042.51+323249.0  & 0.034 & -21.63 &  43 & 28 & 17 &   3 &  -4.76 &  736 & -538\\
SDSS~J133455.94+134431.7  & 0.023 & -22.16 &  14 &  3 & 14 &  20 &  -4.07 &  222 & -196\\
SDSS~J133621.18+031951.0  & 0.023 & -21.74 &  12 &  7 & 10 &  12 &  -4.45 &  296 &  133\\
SDSS~J133737.49+155820.0  & 0.026 & -22.32 &   6 &  3 &  6 &  10 &  -4.47 &  205 &  257\\
SDSS~J134159.72+294653.5  & 0.045 & -22.05 &  52 & 44 & 29 &  53 &  -4.82 &  462 &  632\\
SDSS~J135036.01+334217.3  & 0.014 & -21.40 &   7 &  3 &  7 &   8 &  -4.61 &  198 & -424\\
SDSS~J135226.71+140528.5  & 0.023 & -22.05 &  12 &  6 & 12 &  16 &  -4.56 &  221 &  -45\\
SDSS~J140528.32+304602.0  & 0.025 & -21.04 &   2 &  1 &  0 &   6 &  -4.97 &  238 & -739\\
SDSS~J141451.35+030751.2  & 0.025 & -22.18 &  20 &  8 & 15 &  16 &  -4.54 &  280 & -326\\
SDSS~J141517.98-022641.0  & 0.047 & -22.42 &   8 &  6 &  2 &   2 &  -5.30 &  850 & -285\\
SDSS~J142724.23+372817.0  & 0.032 & -22.03 &  22 & 14 & 17 &  16 &  -4.15 &   98 &  -53\\
SDSS~J143156.59+164615.4  & 0.048 & -22.70 &  31 & 30 & 11 &   2 &  -4.68 &  331 & -691\\
SDSS~J143312.96+525747.3  & 0.047 & -21.53 &  46 & 38 & 27 &  51 &  -4.92 &   86 &  548\\
SDSS~J143424.79+024756.2  & 0.028 & -21.35 &  28 & 12 & 19 &   8 &  -4.69 &  411 & -103\\
SDSS~J143620.38+051951.5  & 0.029 & -22.19 &   6 &  2 &  3 &   4 &  -4.73 &  165 &  496\\
SDSS~J144745.52+132032.2  & 0.044 & -21.33 &   7 &  7 &  2 &   2 &  -5.14 &  672 & -719\\
SDSS~J145216.49+121711.5  & 0.031 & -21.46 &  10 &  3 &  7 &  10 &  -4.86 &  380 &  218\\
SDSS~J145243.25+165413.4  & 0.046 & -22.56 & 120 &111 & 70 & 167 &  -4.81 &  301 &  289\\
SDSS~J145616.20+203120.6  & 0.045 & -22.59 &   8 &  7 &  3 &   3 &  -4.97 &  360 &  188\\
SDSS~J150152.30+174228.2  & 0.047 & -22.20 &  17 & 14 &  8 &   7 &  -4.54 &  450 &  -69\\
SDSS~J150425.68+074929.7  & 0.049 & -21.72 &  15 & 13 &  5 &   4 &  -4.85 &  473 &  436\\
SDSS~J150601.89+084723.2  & 0.046 & -22.33 &   3 &  3 &  3 &   6 &  -5.81 &  578 &   87\\
SDSS~J150636.57+092618.3  & 0.028 & -21.12 &   5 &  3 &  4 &   6 &  -4.88 &  394 & -154\\
SDSS~J150808.25+265457.6  & 0.033 & -20.63 &   4 &  1 &  3 &   3 &  -4.58 &  304 &  310\\
SDSS~J152010.94+254319.3  & 0.034 & -22.13 &  32 & 21 & 24 &  41 &  -4.44 &  103 &  179\\
SDSS~J152151.85+074231.7  & 0.044 & -22.61 &  81 & 73 & 42 &  89 &  -4.38 &   79 & -197\\
SDSS~J153016.15+270551.0  & 0.033 & -21.51 &  13 &  8 &  9 &  12 &  -4.68 &  433 &  161\\
SDSS~J154147.28+453321.7  & 0.037 & -21.98 &  24 & 15 &  8 &   8 &  -5.41 &  361 &   -2\\
SDSS~J154426.93+470024.2  & 0.038 & -22.49 &  11 &  8 &  6 &   8 &  -4.81 &  335 &  259\\
SDSS~J154451.23+433050.6  & 0.037 & -22.43 &  15 & 13 &  7 &   2 &  -5.34 &  458 &  271\\
SDSS~J155951.61+255626.3  & 0.045 & -21.99 &  10 &  7 &  5 &   2 &  -5.25 &  127 &   49\\
SDSS~J155953.99+444232.4  & 0.042 & -21.88 &  10 & 10 &  3 &   6 &  -4.78 &  361 &  259\\
SDSS~J160426.51+174431.1  & 0.041 & -20.89 &  80 & 63 & 34 &   2 &  -4.68 &  623 &  686\\
SDSS~J160523.84+143851.6  & 0.041 & -22.60 &  25 & 18 &  8 &  21 &  -5.19 &  202 &  565\\
SDSS~J160641.83+084436.8  & 0.047 & -22.16 &   9 &  5 &  6 &   4 &  -4.72 &  766 &   -9\\
SDSS~J161238.84+293836.9  & 0.032 & -21.71 &  23 & 16 & 17 &  34 &  -4.57 &  380 &   84\\
SDSS~J161256.85+095201.5  & 0.017 & -21.49 &   2 &  0 &  2 &   3 &  -4.24 &  235 &  108\\
SDSS~J162146.06+254914.4  & 0.048 & -22.49 &  13 & 10 & 11 &  14 &  -4.67 &  140 &   55\\
SDSS~J162846.13+252940.9  & 0.040 & -21.97 &  25 & 19 & 19 &  32 &  -4.74 &  227 & -360\\
SDSS~J162944.98+404841.6  & 0.029 & -18.99 & 127 & 82 & 92 &   3 &  -4.08 &  222 & -385\\
SDSS~J164925.86+360321.3  & 0.032 & -21.63 &   8 &  5 &  5 &   7 &  -4.62 &  369 & -164\\
SDSS~J165830.05+252324.9  & 0.033 & -21.49 &   3 &  1 &  3 &   2 &  -4.74 &  228 & -713\\
SDSS~J170358.49+241039.5  & 0.031 & -22.31 &   2 &  2 &  2 &   6 &  -4.78 &  119 &  175\\
SDSS~J171522.97+572440.2  & 0.027 & -22.81 &  67 & 39 & 52 & --- &  -4.10 &  111 & -535\\
SDSS~J172215.41+304239.8  & 0.046 & -22.87 &  25 & 22 & 19 &  39 &  -4.54 &  394 &  -92\\
\hline
\hline
\end{longtable}
\end{center}
Column description: (1) source name; (2) redshift; (3) SDSS DR7 r band
AB absolute magnitude; (4 and 5) number of cosmological friends within
2 Mpc, observed and simulated at z=0.05, respectively; (6) number of
cosmological friends within 1 Mpc; (7) N$_{\rm gal}$ from T12; (8)
fifth nearest neighbor density $\Sigma_5$; (9 and 10) projected
distance in kpc (\Rcn) and redshift difference (times the speed of
light) in \kms (\vpar) from the average of the cosmological neighbors.
\twocolumn

\onecolumn
\begin{center}
\begin{longtable}{l r r r r r r r r r}

\caption[Properties of the \sFR\ 
  sources. ]{Properties of the \sFR\ 
  sources.}

\label{tabs} \\

\hline \hline 
name &  z & M$_{\rm r}$ & \Ndue &  $N_{\rm cn,0.05}^{2000}$ & \Nuno & N$_{\rm gal}$ & $\log \Sigma_5$ & \Rcn & \vpar \\
\hline	
\endfirsthead

\multicolumn{3}{c}{{\tablename} \thetable{} -- Continued} \\[0.5ex]
\hline \hline 
name &  z & M$_{\rm r}$ & \Ndue &  $N_{\rm cn,0.05}^{2000}$ & \Nuno & N$_{\rm gal}$ & $\log \Sigma_5$ & \Rcn & \vpar \\
\hline
\endhead

\hline
  \multicolumn{10}{c}{{Continued on Next Page}} \\
\endfoot

  \\[-1.8ex] 
\endlastfoot

SDSS~J090100.09+103701.7 & 0.029 & -22.54 &  11 &   2 &  3 &   2 &  -5.38 &  789 & -343\\
SDSS~J092122.11+545153.9 & 0.045 & -22.35 &  60 &  47 & 27 &  55 &  -4.77 &  217 & -193\\
SDSS~J092151.48+332406.5 & 0.024 & -22.12 &  78 &  21 & 43 &   4 &  -4.45 &  469 &  266\\
SDSS~J093957.34+164712.8 & 0.047 & -21.71 &  12 &  11 &  7 &  10 &  -4.19 &  398 & -597\\
SDSS~J101623.01+601405.6 & 0.031 & -22.55 &  16 &  10 & 15 &  17 &  -4.82 &   63 &  -13\\
SDSS~J104740.48+385553.6 & 0.035 & -22.84 &  46 &  24 & 22 &  33 &  -4.38 &  489 &    9\\
SDSS~J111125.21+265748.9 & 0.034 & -22.61 &  63 &  16 & 16 &  21 &  -3.85 &  425 & -418\\
SDSS~J132451.44+362242.7 & 0.017 & -21.83 &  20 &   4 & 10 &  10 &  -4.17 &  499 & -469\\
SDSS~J133242.54+071938.1 & 0.023 & -22.15 &  46 &  12 & 21 &  30 &  -4.05 &  447 &  -39\\
SDSS~J145222.83+170717.8 & 0.045 & -22.32 & 113 & 100 & 37 &   2 &  -4.81 &  691 &   46\\
SDSS~J155603.90+242652.9 & 0.043 & -22.44 &  18 &  10 &  9 &   4 &  -4.69 &  496 &  181\\
SDSS~J155749.61+161836.6 & 0.037 & -23.12 &  44 &  26 & 26 &  31 &  -4.80 &  538 &   21\\
SDSS~J160332.08+171155.2 & 0.034 & -22.46 & 164 &  38 & 44 &   7 &  -4.16 &  424 & -259\\
SDSS~J160722.95+135316.4 & 0.034 & -22.39 &  49 &  16 & 24 &  24 &  -4.49 &   87 & -294\\
\hline
\hline
\end{longtable}
\end{center}


\begin{center}
\begin{longtable}{l r r r r r r r r r}

\caption[Properties of the \FR\ sources with $z<0.05$. ]{Properties of the \FR\ sources with $z<0.05$.}
\label{tabf} \\

\hline \hline 
name &  z & M$_{\rm r}$ & \Ndue &  $N_{\rm cn,0.05}^{2000}$& \Nuno & N$_{\rm gal}$ & $\log \Sigma_5$ & \Rcn & \vpar \\
\hline	
\endfirsthead

\multicolumn{3}{c}{{\tablename} \thetable{} -- Continued} \\[0.5ex]
\hline \hline 
name &  z & M$_{\rm r}$ & \Ndue &  $N_{\rm cn,0.05}^{2000}$& \Nuno & N$_{\rm gal}$ & $\log \Sigma_5$ & \Rcn & \vpar \\
\hline
\endhead

\hline
  \multicolumn{10}{c}{{Continued on Next Page}} \\
\endfoot

  \\[-1.8ex] 
\endlastfoot

SDSS~J100451.83+543404.3   & 0.047 & -22.63 &  44 & 39 &  17 &  43 &  -4.64 &  197 &  -55\\
SDSS~J103258.88+564453.2   & 0.045 & -22.96 &  55 & 45 &  36 &  65 &  -4.09 &  229 & -109\\
SDSS~J104921.13$-$004005.0 & 0.039 & -22.65 &  13 &  9 &   8 &   2 &  -5.23 &  173 &  250\\
SDSS~J113359.23+490343.4   & 0.032 & -22.62 &  86 & 46 &  60 & 105 &  -4.00 &  238 & -212\\
SDSS~J120401.47+201356.3   & 0.024 & -22.11 &  84 & 31 &  70 & 152 &  -4.00 &  164 &  192\\
SDSS~J141652.94+104826.7   & 0.025 & -23.14 &  38 & 15 &  31 &  46 &  -4.05 &  163 &   58\\
SDSS~J145555.27+115141.4   & 0.032 & -22.52 &  17 &  8 &   9 &  19 &  -4.60 &  490 &  304\\
SDSS~J155721.38+544015.9   & 0.047 & -22.68 &  16 & 12 &   9 &  12 &  -4.20 &   92 &   36\\
SDSS~J161114.11+265524.2   & 0.032 & -22.31 &   8 &  4 &   6 &  15 &  -4.68 &  189 &  -49\\
\hline
\hline
\end{longtable}
\end{center}
Column description: (1) source name; (2) redshift; (3) SDSS DR7 r band
AB absolute magnitude; (4 and 5) number of cosmological friends within
2 Mpc, observed and simulated at z=0.05, respectively; (6) number of
cosmological friends within 1 Mpc; (7) N$_{\rm gal}$ from T12; (8)
fifth nearest neighbor density $\Sigma_5$; (9 and 10) projected
distance in kpc (\Rcn) and redshift difference (times the speed of
light) in \kms (\vpar) from the average of the cosmological neighbors.
\twocolumn

\bibliographystyle{./aa}

\end{document}